# Structured Transparency: a framework for addressing use/mis-use trade-offs when sharing information

Andrew Trask[§‡✳] Emma Bluemke[†§‡] Teddy Collins[§‡] Ben Garfinkel[§‡] Eric Drexler[‡]

Claudia Ghezzou Cuervas-Mons[§] Iason Gabriel[✳] Allan Dafoe[‡✳] William Isaac[‡✳]

## ABSTRACT

Successful collaboration involves sharing information. However, parties may disagree on how the information they need to share should be used. We argue that many of these concerns reduce to ‘the copy problem’: once a bit of information is copied and shared, the sender can no longer control how the recipient uses it. From the perspective of each collaborator, this presents a dilemma that can inhibit collaboration. The copy problem is often amplified by three related problems which we term the bundling, edit, and recursive enforcement problems. We find that while the copy problem is not solvable, aspects of these amplifying problems have been addressed in a variety of disconnected fields. We observe that combining these efforts could improve the governability of information flows and thereby incentivise collaboration. We propose a five-part framework which groups these efforts into specific capabilities and offers a foundation for their integration into an overarching vision we call “structured transparency”. We conclude by surveying an array of use-cases that illustrate the structured transparency principles and their related capabilities.

## 1 Introduction

Collaboration requires sharing information amongst participants. This gives rise to the central problem of information governance, which we call the *copy problem*: after replicating and sharing a bit of information, the sender can no longer control how the recipient might use it. The copy problem creates a challenging trade-off for would-be collaborators because each participant must make an educated guess about the likelihood and impact of others’ misuse of the shared data. Assuming rationality, each participant weighs the expected cost of this process against the expected benefit of the collaboration. If the latter outweighs the former for all parties, then the collaboration proceeds. However, if a would-be collaborator expects costs to exceed rewards, they may limit or block the collaboration. Relevant risks and costs informing this assessment can include privacy, security, legal, IP, competitive, public relations, and other similar considerations.

The first contribution of this paper is to lay out the copy problem, which poses serious issues because it cannot presently be solved (information cannot be controlled once shared) and because it scales quadratically in the size of a collaboration (*n* parties must satisfy *n(n-1)* data-sharing relationships to proceed)[1]. Since this dilemma applies anywhere information would be shared in a collaboration, it has a broad impact.

The second contribution of this paper is to argue that three additional issues, which we term the bundling, edit, and recursive enforcement problems, exacerbate the copy problem. In many cases, these challenges can be solved, reducing the impact of the copy problem. However, we observe that progress on these issues is fragmented across many disciplines, such as machine learning, cryptography, distributed systems, database theory, statistics, probability, political science, and legal studies.

As the third contribution, this work re-frames these related-but-disconnected approaches as a shared aspiration for *structured transparency* (e.g., the ability for actors to reduce collaboration risks and costs by defining and enforcing precise flows of information) and organises techniques under a five-part framework. This helps point out which approaches offer identical capabilities and which are complementary to one another. We hope that this framework will serve as a foundation for integrating solutions to the edit, bundling, and recursive enforcement problems, and, in turn, dull the effect of the copy problem and decrease the informational cost of collaborations.

---

[1] One analog means to satisfy n(n-1) relationships might be to select a subset of the parties to be trusted to craft an information flow on behalf of the others. If just one party is selected they are the “trusted third party”, if several they are a committee.

† Institute of Biomedical Engineering, Department of Engineering Sciences, University of Oxford
‡ Centre for the Governance of AI, University of Oxford.
§ OpenMined
✳ DeepMind

The fourth contribution of this work describes technical approaches for structured transparency, especially a set of techniques known as *privacy enhancing technologies* (PETs). We find that, when viewed through the lens of structured transparency, PETs offer enhancements far beyond that of privacy, addressing the edit, bundling, and recursive enforcement problems more generally — with relevance for many of the factors that weigh on collaboration decisions (security, legal, IP, competition, public relations, etc.). Finally, we illustrate the contributions of PETs within the structured transparency principles via a set of real-world use cases.

## 2 The Copy Problem and Ideal Information Flows

We begin our framing of the copy problem by considering a framework proposed by Nissenbam in her seminal work "Privacy in Context" [54]. In this work, she proposes that privacy is chiefly concerned with "information flows in context", such that compromising privacy is not as simple as a specific category of data (sensitive vs non-sensitive data) or a specific set of actors (governments surveilling citizens, etc.). Nissenbaum argues that privacy is about holistically understanding how multiple parties are participating in a flow of information and the degree to which that collaboration is acceptable according to an appropriate moral framework.

We generalise this idea of an information flow in context slightly, broadening it to include other concerns (beyond privacy) that might lead to informational risks and costs relating to: intellectual property, security, legal restrictions, competitive dynamics, safety, etc. However, the fundamental framing of an information flow in context remains the same.

Building on this premise, we introduce the concept of an *ideal information flow*, which is a flow of information that maximises fulfilment of informational exchange necessary for a collaboration while minimising the possibility of any problematic retention or use of said information thereafter[2]. An information flow conforms to this ideal under the assumption that there are no constraints on the degree of specificity an information flow could entail. For example, for a patient going to a doctor, the ideal information flow may allow a patient to receive a treatment with a high degree of confidence on validity of that recommendation while never revealing any information about their medical status in the process. This theoretical aim is "ideal" in that it answers the principal question of a doctor-patient interaction ("How do I get healthy?") without answering any other questions. The doctor could presumably bill for their services, the patient would recover, and no-one would learn any other information. As a nuance, in a truly ideal information flow, the patient wouldn't even be able to reverse-engineer information about the doctor or their practice from the diagnosis they received. The information flow would deliver purely and exclusively the desired answer and nothing more, either directly or indirectly.

The purpose of the structured transparency framework is to frame a set of tools which enable progress towards this ideal relative to analog or legacy approaches to collaboration. A flow that does not achieve this ideal conveys non-essential information that we refer to as "collateral information leakage". As such, another way of describing the aim of structured transparency is to facilitate a collaboration's necessary or useful information flows without collateral information leakage between collaborators.

**Common informational risks and costs**

While a full survey of collaborations mitigated by the copy problem is out of scope, we briefly survey several notable themes. Such issues, highlighted via a selection of profiles below, illustrate the expansive issue of collaborators who cannot work out their disagreements about how their information should be used.

Consider an actor who sells information for a profit. Each customer obtains the ability to sell that information further, thereby becoming a potential competitor. As a result, often the most profitable way to leverage data is to simply exploit it for internal use, and not share it at all. For example, a healthcare firm might train AI models to detect disease using only their own data but decline to contribute to the AI models of other similar hospitals in their area. However, this leaves significant societal utility on the table, especially given the valuable insights that datasets can produce when combined.

An analogous market failure occurs within academia, wherein academics have a strong incentive to avoid sharing research data until they have fully exploited its research opportunities. The cost is slower scientific progress, as researchers are limited to data available at their own academic institution instead of being able to leverage research data across up to (perhaps) thousands of other institutions. This has practical as well as academic costs: by inhibiting advances in domains such as medicine and climate science, barriers to information sharing can leave society at large worse off. Particularly within fields such as healthcare, the market failure is profound to consider.

[2] Given the multi-objective nature of satisfying the desires of multiple collaborating parties, this can be thought of as a theoretical Pareto-optimal point which the multi-objective pareto frontier desires to achieve.

**Complicating factors**

A complicating factor in such collaborations can be a power imbalance between would-be collaborators. A regulator collaborating with a market participant to audit their operations may not offer the market participant an option to not collaborate. In this case, no matter how high the perceived costs of collaboration could be to the market participant, those costs must be paid. Similarly, an individual collaborating with a digital service provider (by using their online service) to socialise with their friends might not wish to reveal their personal interactions to a for-profit company, yet the online service might not choose to offer this option.

## Information Flow Taxonomy

Before describing the tenets of structured transparency, we outline three broad categories of information flows in ascending order of complexity. These categories provide examples of the diverse situations where structured transparency can be applied and serve as useful archetypes that we will refer to throughout the paper.

**Messaging flows** seek to transfer bits of information from one party to another such that they can be trusted by the receiver (verified) and do not reveal other bits to the receiver. Such transfers can face challenges in practice, but existing social institutions possess useful techniques for enabling trustworthy messaging flows. In the analog world, a sealed envelope enables the postal service to deliver a message without inspecting its contents. In the digital world, encrypted messaging platforms like Signal and WhatsApp perform a similar function.

**Service provider flows** are messaging flows that involve computing on the message during transit. An intermediary (service provider) receives bits in an information flow, uses them for computation, and then forwards the result to the next party in the information flow. Using legacy approaches, the service provider is a trusted third-party who must be able to view all of the inputs to the computation in order to generate the correct output. Consider, for example, situations – analog or digital – in which a service requires personalization, such as a doctor's visit: a patient gives information to a doctor, the doctor uses that information to choose a diagnosis and treatment, providing this information to the patient. Similarly, a film recommendation service receives information from a user relevant to film preferences, infers what new films they might enjoy, and then provides a ranked list to the user. Customised services like these almost always require sharing personal data, which creates "keep a copy" risks.

Finally, **aggregation flows** are service provider flows in which the service provider receives and combines messages from multiple parties. Aggregation flows contain all the challenges of service-provider flows with the additional requirement that the aggregating entity itself intends to produce a message from some aggregation, rather than simply processing and forwarding information. This could be a high-level trend, such as the number of COVID cases per day, or a needle-in-a-haystack problem that involves pinpointing a specific fact or individual within a large group.

## 3 The Components of Structured Transparency

Providing effective structured transparency requires fulfilling five main criteria: input privacy, output privacy, input verification, output verification, and flow governance. Not every situation requires that all be explicitly addressed, but most copy-problem-related obstacles to collaboration can be reduced to some combination of these issues. Below, we describe each of these from the perspective of someone participating in an information flow.

**Input privacy** refers to the ability to process information that is hidden from you or, symmetrically, to allow others to process your information without revealing it to them[3][4]. Consider the sealed envelope again: this allows information (the contents of a letter) to be processed (transmitted) by a mail service without that intermediary gaining access to it. The envelope provides input privacy.

**Output privacy** allows you to receive/read the output of an information flow without being able to infer further information about the input or, symmetrically, to contribute to the input of an information flow without worrying that the subsequent output could be reverse engineered to learn about your input [12,42].

While the aim of output privacy is similar to that of input privacy (both relate to protecting providers of inputs to an information flow), they are not the same. Input privacy is concerned with *facilitating the flow of information* without leaking collateral data, whereas output privacy is concerned with *preventing the output of the flow from being reverse engineered* to reveal additional information about the input. Input privacy is concerned with preventing parties participating in *computation* from learning anything about each other's inputs[5], whereas output privacy is concerned with preventing the *recipient* from inferring the inputs. Returning to the sealed letter example, output privacy in this context would guarantee that the letter recipient could not deduce sensitive information that the author did not want to include in the message[6].

---

[3] More formally, input privacy is satisfied when multiple data-holding parties can provide inputs to one or more computing parties, who can provide computation without having the ability to know the values of the respective inputs, intermediate variables, or outputs.

[4] We also consider "semi-input privacy" to be when a group of parties compute a flow together where only some successfully hide their inputs to the flow (such as vanilla federated learning, as we will see)

At first glance, satisfying input and output privacy may appear sufficient to enforce any information flow. However, most use cases require balancing these criteria with the informational burdens necessary for recipients to trust the bits they receive. Privacy, in other words, exists in mild tension with verification.

This tension arises because often actors use *internal consistency* to determine whether an information flow is valid. Consider a barkeep viewing a driver's licence to check whether someone is of drinking age. They are consuming an information flow that began with a birth certificate, continued through to a regional transportation authority, and terminated in the contents of a driver's licence. A barkeep determines whether a licence is valid based on all of the attributes of the licence appearing coherent and consistent with the issuer's standards. If, for example, the owner of that driver's licence used scissors to cut out the name and home address on their driver's licence (to preserve privacy), the document would no longer be valid and the barkeep may deny them alcohol. This is true even if the licence still contained a photo and birthdate. As such, verification is often a source of collateral information leakage, which other techniques can help to alleviate.

**Input verification** allows you to verify that information you receive from an information flow is sourced from entities you trust, or, symmetrically, it allows you to send information such that the output can be verifiably associated to you. When you sign a document, you make a mark which (in theory) only you can make – a signal to readers of the document that you approved the information contained therein. Novel input verification techniques empower a signer to verify *specific* attributes of an input to an information flow, such as that it came from a trusted source or that it happened within a specific date range [44]. Often this is framed as a "certificate", "verification", or "attestation", wherein an input party has signed that a specific fact being conveyed is something they believe to be true. (Note that this in no way guarantees a statement to actually be true, merely that a party claims it to be so.)

**Output verification** allows you to verify attributes of any information processing (computation) within an information flow. In analog systems that contain little to no computation, the difference between input and output verification can be nuanced or nonexistent. Process auditing by an external party, however, is an example of output verification. If a letter goes missing, one might audit tracking records. Likewise, a tax auditor might verify that the flow of information (in this case, funds) from a company's account to their employees' accounts obeys relevant tax codes.

Finally, **flow governance** is satisfied if each party with concern/standing over how information should be used has guarantees that the information flow will adhere to their intended uses. This is important because even if a flow satisfies the necessary criteria of input and output privacy and input and output verification, questions still remain concerning who holds the authority to *modify* the flow [55]. In real-world scenarios, a wide variety of governance mechanisms exist for this purpose (escrow and executorship stand out as widely-used legal devices). Within the physical world, multi-key safety-deposit boxes for holding secure documents accomplish similar goals.

To recap these characteristics in the context of sending a letter, we can describe input privacy as the protective envelope that prevents unauthorised access, while output privacy involves withholding sensitive personal information from the letter. Input verification can be likened to the signature on the letter, while output verification corresponds to a wax seal that assures the recipient that the letter has not been tampered with. Finally, flow governance corresponds to shipping the letter in a secure safe with a combination lock, where only a set of governing parties (such as the sender or receiver) knows the correct combination to the safe. From a structured transparency perspective, when an information flow possesses all of these guarantees to a sufficient degree, we consider the flow to be a *well formed information flow* — capable of delivering a specific, known set of bits from a set of input parties to a specific set of output parties with verification and without collateral information leakage.

## 4 Copy Problem Amplifications as Limitations to Structured Transparency

Satisfying all the criteria of structured transparency can prove difficult in practice. As mentioned, for example, input and output verification can come at the cost of input and output privacy. The central issue is the copy problem, alongside three related problems which amplify it: the bundling problem, the edit problem, and the recursive oversight problem.

[5] Except the output destined for them (if any).
[6] For example, an anonymous author may wish to send a message without inadvertently revealing too many facts such that the identity of the sender can be inferred by the recipient using context clues.

**The copy problem** lies at the root of many structured transparency challenges. When a bit of information is shared, the recipient gains control over its use, and they are generally not constrained by any technical limitations that would prevent them from misusing it (although institutional rules and norms may act as constraints, and legal or social repercussions may exist after the fact). As a result, when deciding whether to share information, data owners often face a trade-off that pits the benefits of sharing with the risks of misuse.

**The bundling problem** amplifies the copy problem. It is often difficult to share a bit of information without also needing to reveal additional bits because either the conventional encoding will not allow individual bits to be shared or a bit cannot be trusted/verified without the context of other relevant bits. Take, for instance, the use of a surveillance video: many pieces of irrelevant (and potentially invasive) information are shared to contextualise the critical piece(s) of information (e.g. whether a suspect was in a particular location). Another example is a driver's licence, which reveals all of the details on the card in order to verify a single piece of information, namely whether the individual is old enough to enter a given venue. Put another way, while cutting out one's birthday from a licence would effectively hide the other information on the card, it would not suffice to enter an age-restricted establishment.

**The edit problem** also amplifies consequences of the copy problem. If an entity that stores a piece of information makes an edit before transmitting it to another party, the recipient has no way of knowing that the information was altered. This is similar to the phenomenon observed in the children's game of telephone. Relatedly, a bank balance is stored by the bank itself (as a trusted third-party), rather than by the account holder who might be inclined to make edits to it.

The use of third-party oversight institutions can solve issues caused by the copy, bundling, and edit problems. In doing so, however, this solution presents a fourth issue: the **recursive oversight problem**. When one party oversees the use of information, it creates another, even more knowledgeable entity that could potentially misuse the information. This raises the question of "who watches the watchers?" – in other words, how can we ensure that the oversight institution itself is trustworthy and accountable?

In practice, these three hurdles amplify the copy problem and further constrain the ability for collaborators to satisfy the guarantees of structured transparency in a wide variety of contexts. This results in reduced collaborations and a higher degree of perceived cost for collaborators who do proceed. Without new capabilities that could enable more precise information flows, collaborators will continue to be constrained by trade-offs related to the copy problem.

## 4 Technical Tools for Structured Transparency

Fortunately, many information flows constrained by the edit, bundling, and recursive enforcement problems can be addressed elegantly in the digital domain. And while the copy problem may never truly be solved (and arguably should not be [11]), combining relevant technical tools with social and legal measures can effectively reduce the potential harm it poses — in some cases fully achieving the ideal information flow.

Such technical tools can be loosely grouped in relation to the five sub-problems of structured transparency outlined in Section 3.

### Technical Input Privacy

All proposed within the last half-century, technical input privacy tools come primarily from the field of cryptography: public-key cryptography, end-to-end encryption, secure multi-party computation, homomorphic encryption, functional encryption, garbled-circuits, oblivious RAM, federated learning, on-device analysis, and secure enclaves are several popular (and overlapping) techniques capable of providing varying degrees of input privacy [2,3,6,8,14–16,21,23,24,34,40–42]. Some of these techniques can theoretically facilitate any arbitrary computation (also known as 'Turing-complete computation') while keeping the computation's inputs secret from all parties involved[7]. These methods differ in terms of performance, cost, trust model, type of computation they are most suited for, etc. For example, homomorphic encryption (HE) often requires heavy computation even for relatively simple information flows, while secure multi-party computation requires less computation than HE but greater message volume between the various parties in the flow (increased network overhead) [42]. Perhaps the most performant input privacy technique occurs via secure enclaves, which require the use of specialised hardware. Of particular note are GPU enclaves, which offer near-equal performance to top-of-the-line GPUs and are expected to be generally available in the cloud in the near future[8]. Input privacy techniques still lack general-purpose software necessary for widespread use, but this is an active and quickly-maturing area of software engineering [56].

[7] We consider techniques which only keep some parties' inputs hidden from other participants to be "semi-input privacy"

The most important implication of technical tools for input privacy is that they can, theoretically, perform a task seemingly impossible in the analog world: achieve service provider information flows without a trusted third-party [42]. In other words, technical input privacy tools could allow service providers to process data without being able to see it or use it for other purposes outside of the governed information flow. This means that claims along the lines of "we need a copy of the data in order to provide a service that takes it as input" will lose legitimacy as technical input privacy tools mature.

Aggregation flows also stand to benefit from input privacy tools. Under the right configuration, the aggregator will learn only an output intended for them, and will be unable to observe the computation inputs, or any outputs not allowed by those governing the flow.

## Technical Output Privacy

Early forms of non-technical output privacy focused on redacting sensitive data-points or threatening reprisal via legal or other means against those who reverse engineer inputs. For highly bundled data, redaction can prove very challenging, and even when one can redact information easily (such as removing names and addresses from a database), the advent of big data has demonstrated that 'anonymized' data can sometimes be de-anonymized if enough latent signal is available [25,27,28,31,32,36,37].

However, technical output privacy tools (chiefly, differential privacy and related techniques) can provide strict upper bounds on the likelihood that a data point could be reverse-engineered to convey information about a specific party in the input data [12,13]. This capability is useful in many settings, but it has particular significance in aggregator flows where the actor processing the information is performing statistical analysis; with differential privacy, aggregator flows can reveal high-level insights without ever disclosing individuals' data in detail. This holds promise for preserving privacy in the context of scientific inquiry, official statistics, and particular use cases of surveillance (such as public-health surveillance used to track the progression of COVID-19).

If one splits a set of bits into sections, differential privacy provides a limit on the likelihood that a statistician who subsequently aggregates these bits could learn information about any specific section of bits. For instance, when applied at the level of individuals, these techniques protect privacy by preventing one from learning information about any particular person. However, if data was split by company, nation, or other attribute for which linkability would like to be prevented, differential privacy is a useful tool in these settings as well. For example, if documents were grouped by the author's employer, and then an aggregate statistic was computed across documents, differential privacy could be used to prevent one from knowing information specific to an employer.

## Technical Input Verification

Within the context of message flows, some input verification techniques such as cryptographic signatures are robust, performant and proven at scale, but advanced techniques (like zero-knowledge proofs) are still relatively nascent owing to a lack of awareness and early-stage tooling among the PET community [56-57]. Most input verification techniques use some combination of public-key infrastructure (Trust-over-IP/SSI, Key Transparency, etc.), cryptographic signatures, input privacy techniques with active security, and/or zero-knowledge proofs [14,17,22]. These methods can allow an actor to verify a specific attribute such that the information flow output contains cryptographic proof of this verification.

For example, consider the driver's licence example mentioned above: normally, somebody inspecting a driver's licence views the card in its entirety. Technical input verification tools do not suffer from this constraint [35], since they can verify and reveal individual attributes within an information flow. This allows for high levels of both input privacy and input verification, effectively eliminating one of the trade-offs inherent in achieving structured transparency in analog settings.

As with input privacy, however, the greatest promise of input verification techniques lies in the ability to verify the inputs to arbitrary computations. A simple example of this is verified voting, wherein an election authority can publish the final vote with cryptographic evidence guaranteeing to each voter that their vote was included in the final tally without violating the privacy of a secret ballot [58]. Because technical input verification techniques make it possible to verify properties of specific inputs selectively, it enables systems which, in practice, can be both more private and more strongly verified than analog equivalents.

[8] https://www.nvidia.com/en-us/data-center/solutions/confidential-computing/

## Technical Output Verification

The key limitation of output verification tools in many analog contexts is that the verifier must examine the data in order to perform the verification. For example, a tax auditor must inspect cash flows in detail in order to determine whether fraud has occurred, and a financial regulator must have access to credit score inputs and outputs in order to ascertain whether loans have been distributed fairly. This in turn relates to the recursive oversight problem: effective oversight requires granting access that opens the door to more potential misuse.

However, when combined with the aforementioned input privacy techniques, technical tools for output verification can address this challenge. An external auditor could verify properties of an information flow without learning anything beyond the output of targeted tests (e.g. searching for patterns reflective of fraud) while also ensuring that the tests ran correctly. Such capabilities could increase the precision, scale, and security of auditing institutions, potentially facilitating new types of checks and balances and fairer distributions of power. In addition, output verification relates to ongoing research for auditing or evaluating models for fairness, bias, or emerging dangerous capabilities [5,9,26,30,33,39].

## Technical Flow Governance

As noted above, traditional analog methods of information flow governance rely heavily on legal and physical measures. However, technical tools for flow governance offer distinct advantages in terms of scalability and efficiency over their analog counterparts (see "policy enforcement" in [42]). Secure multi-party computation (SMPC) serves as an excellent example of this: with SMPC, parties can be selected to govern the flow of any given information[9]. Rather than relying solely on legal repercussions for violations, SMPC can implement hard cryptographic limitations to prevent unauthorised behaviour, establishing trust in the system. For example in additive secret sharing (a form of SMPC), a single number can be divided into cryptographic "shares" which, once distributed amongst "shareholders", empower each of those shareholders with veto power over when and how that number (and any subsequent numbers it is used to create) can be used [8]. As all digital computation occurs over numbers, this is a very powerful and general means of enforcing flow governance. Secure enclaves and other input privacy techniques can also offer this ability.

Flow governance can also be enacted over computation [42]. Whereas service provider flows and aggregation flows previously required a trusted third party to observe all inputs to a computation in order to create the output, technical flow governance systems such as SMPC have no such limitation. By performing computations over encrypted, mutually governed information, digital systems can enforce agreed-upon checks and balances among shareholders, in ways that are not possible in physical systems.

## In Combination: A New Frontier for Structured Transparency

Together, these tools can improve governance of information flows and thereby reduce the costs associated with sharing information. They also enable much more precisely structureed social and technical arrangements than their analog predecessors. Most importantly, they provide (1) the ability to unbundle information such that one needs to share only the bits necessary for a collaboration, (2) the ability to ensure (for example, by checking a hash) that data hasn't been edited, (3) a solution to the recursive enforcement problem such that weak actors can audit information about strong actors to which they do not themselves have access (without necessarily becoming strong), and (4), advanced flow governance tools like SMPC which allow actors to share governance over future uses of digital systems even and especially if uncertainty about future uses/mis-uses remains.

While technology alone cannot solve all information-sharing issues, these new capabilities can in combination complement legal and social systems to deliver holistic solutions that were previously impossible. For example, consider a mobile app that collects and stores extensive personal information about its users (location, health data, etc.). Without modern structured transparency tooling, users would have to accept the risks of misuse as the price of using the product in order to receive whatever service it offers (directions, medical advice, etc.). However, with an awareness of these techniques, key members of society could mandate that any personal data remain on-device, or that any off-device operations performed on such data must follow structured transparency principles. Advocacy and activist groups could make such requests, and consumers could expect and demand more of the services offered to them. With knowledge of these technologies, regulators, ethicists, and relevant governance bodies — can push for adherence to more responsible information flows. Note that knowledge is as important as capabilities: Without knowledge of capabilities, we cannot know what to expect, or to demand.

[9] SMPC enables the selection of arbitrary parties to govern the flow over arbitrary information, limited only by the compute and network resources of the chosen information shareholders.

## 5 Illustrations of Structured Transparency

To help ground the framework of Structured Transparency in the context of real-world problems, we survey the following real-world use cases.

### Improved Data Flows for Open Research

For years, the open-data movement in research has expressed concern that the inability to share data safely (primarily due to concerns relating to the copy problem) hampers research progress [1,7,10,18,29,43]. Structured transparency enables information flows that can answer specific research questions while letting the data owners maintain control over the only copy of their research data. For instance:

- Technical input privacy techniques could enable data owners (hospitals, labs, statistics offices, etc.) to grant researchers the ability to perform specific computations over their data without providing access for any other operations. This allows the researcher to answer their research question ('what is the mean weight of newborns in this city?', 'how does my algorithm perform on your clinical MRI dataset?') without the owner facing the perils of the copy problem [19,20].
- Using technical output privacy techniques, data owners could prevent reverse engineering of the computation output to reconstruct inputs, such as a hospital's patient information.
- Input verification techniques could allow data owners to prove to researchers various attributes of the dataset, such as whether or not it was used in another experiment (a potentially important feature for research reproducibility).
- Output verification may also be required in situations that feature a competitive relationship between institutions or research groups. It could, for example, be used to prove that a key statistical result was actually computed by the data owner using the computations requested by the researcher (as opposed to stemming from shortcuts or mistakes).
- Flow governance could distribute control across third parties (e.g. funding bodies, stakeholders in a collaboration network, groups safeguarding rights for vulnerable populations, etc.) to enable especially sensitive information to remain available for appropriate research while minimising risk of misuse.

In summary, stronger, more precise, and more automated controls over data sharing could make more scientific data available for research, increasing the pace of scientific research in many empirically-driven fields.

### Large-Scale Collaboration for Social Good

Addressing many societal challenges will require large-scale coordination. However, privacy concerns can impede such arrangements. To take a specific example, consider energy efficiency: the collection of detailed energy usage data from smart metres has immense potential both to reduce unnecessary carbon emissions and to save consumers money. This data, however, could also be used to infer occupancy and activity patterns in great detail, down to which television channel is being watched [38].

Structured transparency tools could enable the use of insights from usage data to optimise energy consumption, while letting customers maintain control over the only copy of their data. As with the researchers in the example above, the metre company does not need a full copy of this data in order to achieve their goals; they only require the output of specific computations (statistics, model training, or model testing).

Technical input privacy techniques could allow providers to perform their smart service (for instance, turning off lights when no-one is home) using energy usage data without ever seeing it in its entirety. Using technical output privacy techniques, smart-metre companies could prevent any reverse engineering that might infringe customer privacy. Flow governance could distribute control beyond the private companies to third parties such as consumer interest groups unlikely to collude against the consumer in a data attack (e.g. environmental protection or privacy rights activist organisations). In some cases, input or output verification may be necessary. For example, an energy provider might have an incentive to modify the data to give the illusion of complying with a regulation or achieving success in an energy saving campaign. In this way, each aspect of the structured transparency framework has a plausible role in facilitating a well formed information flow.

### Digital Systems of Accountability

Further, we consider the field of accountability holistically. Whether it is a government holding an individual accountable, an individual holding a government accountable, government-to-government, government-to-corporation, or any mixture thereof, accountability is almost universally limited by access to information. Corruption persists when it goes undetected, so why are there not systems offering universal coverage of possible corrupt practices? The answer is plain when considered pragmatically; the relevant information would need to be available for non-colluding parties to consume. However it would be extremely problematic for the email, text, and phone records of citizens, governments, and employees to embrace this form of radical transparency. As a result, accountability and oversight is restrained and some unknown level of corruption persists across all levels of society.

Within the context of structured transparency, “well formed information flows” could conceivably provide for specific bits of information to reach appropriate oversight entities without collateral information leakage — without empowering oversight entities with information other than the specific bits indicating the patterns to which they are responsible for observing. And as a response to the classic “who watches the watchers” challenge earlier, oversight over these oversight entities could also be facilitated via specific flows of information, again without this third party necessarily learning anything about the first party. Future work should explore the implications of accountability under a minimised recursive oversight problem.

## Flow Governance in the AI Lifecycle

At present, AI is experiencing a rapid advancement in its capability and use. As these systems take in larger amounts of information — much of which cannot be manually expected — developing novel infrastructure for governing AI (and its supply chain of data, compute, and talent) is an urgent concern. As such, the creation, deployment, governance, and oversight of AI models is an area of special relevance to structured transparency. As AI systems make their way into increasingly influential roles within society, the number of stakeholders and diversity of incentives increases. This both raises the importance and complicates governance of information flows. Structured transparency principles have many uses across the AI model lifecycle.

**Creation:** We begin by considering governance desires of actors in the supply chain of an AI model. If multiple actors bear the cost of creation by pooling datasets, computational resources, and/or AI research talent in the creation of an AI model, they may wish to ensure that subsequent use of the jointly created asset is governed jointly (for example to distribute profits amongst the group), cryptographically preventing any specific actor from using the model without the other owners’ consent. A situation such as this requires a full structured transparency pipeline, such that actors retain control over their own datasets in the process, and the resulting model resides in a shared governance system such as SMPC or a secure enclave.

**Pre-deployment audit:** After creating a model, the owner(s) may take steps before deploying it, such as subjecting it to various evaluations (accuracy, fairness, bias, robustness, privacy, etc.) [59]. As these models become increasingly influential over outcomes in the real world (consider an AI model driving a car or determining a credit rating for a home-buyer), the efficacy of a model will have increasing importance to users.

**Audit Registration:** We envision a public registry of model audits powered by structured transparency [51]. An auditor who claims to have verified a specific property can upload a signed hash of the model alongside their claim. In other words, auditors could perform evaluations and reliably obtain a unique identifier of the model (its hash) using structured transparency techniques without ever gaining direct access to model weights.

**Deployment and audit verification:** After a model has been deployed, structure transparency tools could verify the predictions it makes to users against an audit registry. Input and output verification techniques can provably provide a prediction alongside the hash of the model that created the prediction and the hash of the inputs that were sent into the model. Thus, for example, someone using a machine translation AI would know that the translation they received used the text they sent and transformed it with a specific model chosen from the audit registry, and that the prediction was the output of only these two inputs.

**Deployment over confidential inputs:** Once a model is deployed, it becomes a part of the information flows of its users. If the model is hosted by its owner, users of the system may be concerned that when they send data to the model for evaluation, the model owner may keep a copy of the data for subsequent use. Input and output privacy and verification can allow users to facilitate an information flow while ensuring that no party sees the data sent to the model. The same information flow could also ensure that nobody but the user sees the model’s predictions either

**Deployment oversight:** While an AI model may have been audited before deployment, the behaviour of an AI model is a function of both the model itself and the data being fed through it. Often one cannot reliably assess a model’s qualities until it is running in the context of its real-world setting [49]. Via structured transparency, an external auditor could evaluate a model in context and produce another claim and hash for the audit registry.

One project is already testing the use of structured transparency tools in this setting, the Christchurch Call Initiative on Algorithmic Outcomes (CCIAO). Announced by New Zealand Prime Minister Jacinda Ardern and French President Emmanuel Macron at a side event to the 2022

United Nations General Assembly, the CCIAO is a joint project between Twitter, Microsoft, and the US and NZ governments [4]. In the coming months, the CCIAO plans to deploy a tool called PySyft to facilitate external civil society researchers using structured transparency principles to audit deployed recommender systems at Twitter and Microsoft. As the Christchurch Call is a coalition including many online service providers, the CCIAO has room for expansion to additional algorithm partners upon successful completion.

This is also highly related to the recently proposed framework of *structured access*, which proposes constrained APIs for overseeing deployed AI systems as an alternative to openly disseminating AI systems[48].

**Deployment co-governance:** Under circumstances where society does not trust the parties that create an AI system to exercise unilateral authority over its operation, structured transparency could allow co-governance by non-colluding actors. For example, an AI firm might elect to share governance with an outside party as an exercise in proving compliance with a norm or law. Under such an arrangement, model use might require unanimous stakeholder approval. This enables oversight that could prove especially valuable in high-impact use cases [50].

**Deployment outcome evaluation:** differing in certain ways from the direct evaluation of an AI model, perhaps the most important metrics for which structured transparency is relevant are real world harms caused by AI. For example, a jurisdiction may wish to ensure that algorithms do not lead to unfair discrimination in the job application process or criminal justice system[45, 46, 47]. While one method might be to evaluate each model individually on a holdout test dataset, such a dataset might not fully capture the risks faced by the model in production. The ground-truth standard for understanding whether an AI model is harmful is the degree to which the deployed system causes harm when deployed. However, data measuring harms is likely to sit across multiple organisations. One can imagine a structured transparency system facilitating a collaboration between organisations that have hiring data (perhaps LinkedIn) and organisations running models these applicants interacted with during their journeys. This could facilitate the answering of questions such as, "For each marginalised and non-marginalized group, what is their relative ability to obtain a computer science job in New York City?". Such a question could take into account the end-to-end outcome for someone living in New York City as opposed to merely whether each individual algorithm is meeting a local standard (e.g., allocating a certain number of hiring recommendations to each group). All the structured transparency guarantees would be essential for the creation of this type of social system, and upon this template the opportunity for ensuring alignment between an AI's impacts and society's values seems ripe for creative exploration.

**Deployment prediction provenance:** Often it is important for someone who receives a piece of information to know whether it was generated from an AI model. When a model is deployed, it can retain a record of all predictions previously made and empower outside users to check whether bits found in the wild were generated by a particular model at some point. While short strings of bits may create false positives, adding additional filters narrowing the list of days, times, and recipients can create additional precision around this type of service.

## 7 Conclusion

This paper attempts to anchor a conversation around structured transparency and its relationship to collaboration. We have outlined a framework that describes how a set of rapidly-developing technical tools can help design and enforce more precise information flows than those currently in widespread use. These methods are not complete solutions in themselves, but by enabling selective sharing and disclosure of information, they offer an opportunity to expand the Pareto frontier of trade-offs between use and mis-use. We have presented several examples illustrating how structured transparency tools could enable collaborators to reduce the informational costs and risks of collaboration. These tools sit at the intersection of a number of active fields of research and have potential applications for many more. It is our hope that this framework acts as a useful bridge between disciplines, and we look forward to receiving feedback.

## ACKNOWLEDGMENTS

This work is the culmination of over two years of both formal research and casual conversation between a wide community of researchers within the AI ethics, safety, governance, and privacy communities around the Centre for Governance of AI. We'd also like to thank Borja Balle, Georgios Kaissis, Jan Leike, Vishal Maini, Kenneth Cukier, Helen Nissenbaum, Phil Blunsom, and Chris Dyer for feedback on ideas and/or early drafts of this work. And finally we'd like to thank OpenMined for providing an environment wherein these social and technical ideas could come together.